\NewCommandCopy{\kernelLabel}{\label}
\documentclass[10pt,aps,pra,superscriptaddress,tightenlines,nofootinbib,showkeys,floatfix]{revtex4-2}
\RenewCommandCopy{\label}{\kernelLabel}
\usepackage[utf8]{inputenc}
\usepackage[T1]{fontenc}
\usepackage{amssymb}
\usepackage{color,graphicx}
\usepackage[caption=false,font=small]{subfig}
\usepackage{lmodern}
\usepackage{microtype}
\usepackage{amsmath}
\usepackage{amsbsy}
\usepackage{bbm}
\usepackage{bm}
\usepackage{epsfig}
\usepackage{braket}
\usepackage{graphics}
\usepackage{appendix}
\usepackage{capt-of}
\usepackage{booktabs}
\usepackage{hyperref}

\begin{document}

\title{Quantum modeling of a radical pair magnetic sensor based on electric dipole moment}

\author{Mahboobe Sehati}
\thanks{These two authors contributed equally to this work.}
\affiliation{Center for Quantum Engineering and Photonics Technology, Sharif University of Technology, Tehran, Iran.}

\author{Ali Soltanmanesh}
\thanks{These two authors contributed equally to this work.}
\affiliation{Center for Quantum Engineering and Photonics Technology, Sharif University of Technology, Tehran, Iran.}

\author{Shabnam Abutalebi}
\affiliation{Center for Quantum Engineering and Photonics Technology, Sharif University of Technology, Tehran, Iran.}

\author{Abolfazl Bahrampour}
\affiliation{Center for Quantum Engineering and Photonics Technology, Sharif University of Technology, Tehran, Iran.}

\author{Naser Haeri}
\affiliation{Center for Quantum Engineering and Photonics Technology, Sharif University of Technology, Tehran, Iran.}

\author{Sareh Rostami}
\email[Corresponding Author:]{sareh.rostami@researcher.sharif.edu}
\affiliation{Center for Quantum Engineering and Photonics Technology, Sharif University of Technology, Tehran, Iran.}

\author{Alireza Bahrampour}
\email[Corresponding Author:]{Bahrampour@sharif.edu}
\affiliation{Center for Quantum Engineering and Photonics Technology, Sharif University of Technology, Tehran, Iran.}
\affiliation{Department of Physics, Sharif University of Technology, Tehran, Iran.}

\begin{abstract}
Photoreduction of cryptochrome in the retina generates magnetically sensitive radical pairs that form the basis of the radical-pair mechanism, a widely studied hypothesis for avian magnetoreception. However, how this magnetic sensitivity is translated into a biologically meaningful signal remains unclear. The absorption of blue light by the flavin adenine dinucleotide (FAD) chromophore can alter the distribution of electrons in cryptochrome and create radical pairs with separated charges. In this study, the spin dynamics of electrons in the radical pair including its spin-orbit coupling were investigated by quantum mechanical modeling. The spin-orbit coupling is weaker than the hyperfine interaction and affects system dynamics only at short inter-radical distances. However, it engages the spatial states of the radical pair and make possible to study spatial related observables. Several interactions were considered in the presence of an external magnetic field, and the resulting electric dipole moment in cryptochrome was computed as the quantity emerging from this coupling. The computations show the induced electric dipole moment clearly depend on the characteristics of the applied magnetic field even after considering dissipative effects. In fact, our findings indicate that the radical pair in cryptochrome protein is a magnetic biosensor, in the sense that in the presence of the geomagnetic field, variations in spin states can influence its electric dipole moment, which may be interpreted via the bird as an orientation signal. The results can be used in the advancement of bio-inspired technologies which replicate animal magnetic sensitivity.
\end{abstract}


\maketitle

\section{Introduction}
The physiological ability to perceive and detect invisible signals of the Earth's magnetic field by an internal and innate compass is the basis of magnetoreception and magnetic navigation in various animals, for example the European robin as a migratory bird. The study to discover and understand this compass has been the subject of extensive interdisciplinary research for decades and continues \cite{wiltschko1975interaction,xu2013estimating,wiltschko2022discovery,schneider2023over,gill2024navigation,kattnig2016chemical}. Since the geomagnetic field is a very weak field that can penetrate biological systems \cite{rankin2015finding} from a biophysical perspective, magnetic sense can be described as the interaction between the magnetic field and biological structure. To date, various experimental studies, including behavioral, histological, and electrophysiological, have been performed on diverse organisms, namely bacteria (magnetobacteria \cite{clites2017identifying, natan2017symbiotic}), invertebrates (such as beetles \cite{vacha2009radio, bazalova2016cryptochrome} and honeybees \cite{liang2016magnetic}), and vertebrates (including birds \cite{wiltschko2021magnetic, wiltschko2019magnetoreception}, fish \cite{formicki2019magnetoreception}, and mice \cite{prato2013magnetoreception, painter2018evidence, malkemper2015magnetoreception}) to investigate magnetoreceptors. In this regard, various theoretical and modeling studies have been conducted to better understand magnetic sense \cite{hiscock2016quantum, adams2018open, mondal2019theoretical, fay2020quantum, babcock2020electron, ramsay2023magnetoreception}. In general, based on these studies, two types of receptors have been proposed to be involved in the mechanism of magnetoreception. One type is an iron-based magnetoreceptor such as biogenic magnetite, which can be studied from a classical perspective. The other type is a magnetosensitive photoreceptor, whose radical pair mechanism is considered as a potential intersection between biology and quantum mechanics. In the mechanism, spin-correlated radical pairs are formed in flavoproteins such as cryptochromes (Cry) during a photon-absorption-dependent process. So far, six Cry types have been identified, with Cry4a and Cry1a being likely candidates for radical pair-based magnetoreception \cite{gortemaker2022direct}. In this context, It is hypothesized that an external magnetic field can affect the spin dynamics of unpaired electrons via the Zeeman effect, which causes splitting and lifting of the degeneracy of triplet sublevels \cite{schulten1978biomagnetic} and changes the interconversion rate between singlet and triplet spin states. Hence, most theoretical studies have only considered spin degrees of freedom in their models \cite{adams2018open, fay2020quantum, sotoodehfar2024quantum}. It should be noted that in addition to its proposed role in magnetoreception, the radical pair mechanism also plays a role in other biological processes, namely reaction centers in photosynthesis \cite{fursman1999distance}, regulation of circadian rhythms in some organisms \cite{yoshii2009cryptochrome}, and DNA repair \cite{doi:10.1021/acscentsci.8b00091}. For a radical pair to act as a magnetic compass, the interconversion between its spin states must be influenced by anisotropic interactions \cite{efimova2008role}. To our knowledge, two main sources of anisotropy are known. One arise from the magnetic coupling between the electron spin and the atomic nuclear spin \cite{bezchastnov2023quantum}, in which isotopic substitution \cite{pazera2023isotope} can change the strength and orientation of the hyperfine coupling and hence modulate the direction-dependent sensitivity, and the other originates from the spin-orbit coupling between the electron spin and its orbital angular momentum \cite{lambert2013radical}.  The radical pair mechanism aligns with the known features of the avian internal compass, including its function as a light-dependent inclination detector \cite{ritz2000model}.  Various pathways \textit{viz-a-viz} chemical and physical transduction, have been proposed to determine the mechanism by which an external magnetic field affects the radical pair, as follows. The change in the interconversion rate of singlet and triplet states induced by the magnetic field, and the subsequent change in the yield of spin states, results in a change in the yield of the chemical product \cite{hore2016radical}. Yoshida et al. modeled biomagnetic chemical signals in the presence of magnetic noise and showed that the compass function of the radical-pair model is robust to intrinsic magnetic noise arising from thermal fluctuations in the electromagnetic field over a realistic range of biomagnetic field strengths \cite{yoshida2025robustness}. It is predicted that the bird can detect this difference in field-dependent yield and use it for navigation. Another biochemical hypothesis supported by experimental evidence suggests that the Cry4a in birds interacts with retinal cone-specific G-proteins, potentially initiating a signaling cascade that alters cell membrane potential \cite{gortemaker2022direct}. In contrast, a biophysical hypothesis proposes that a long-lived triplet radical pair forms in cryptochrome, possessing an electric dipole moment capable of generating an electric field. The field may be able to influence retinal protein isomerization \cite{stoneham2012new, espigule2014physical}. Based on what is briefly reported above, it can be concluded an unambiguous and definite final comment concerning radical pair-based mechanism and signal transduction has yet to be provided, but it seems that the biophysical transduction mechanism offers a simpler way for magnetic field detection.

Given magnetic sensitivity of radical pair system, the aim of this study is to provide, a method for either detecting magnetic fields or transducing them into measurable signals, within the framework of quantum mechanical calculations. For this purpose, the behavior of the system is investigated by considering both the dynamics and transitions related to the spin component and the spatial asymmetry and charge separation related to the spatial component via considering the spin-orbit coupling. Accordingly, the electric dipole moment is expected to emerge as a measurable physical quantity that has the potential to interact with biological structures. In this research, the effect of the characteristics of an external magnetic field on the behavior of the radical pair dipole moment is analyzed in detail. In fact, the considered approach, by integrating spin coherence and electrostatic asymmetry, provides a more complete view in this regard. It is worth noting that in addition to confirming the radical pair as a magnetic biosensor, the results of this study could play a role in the advancement of bio-inspired technologies which replicate animal magnetic sensitivity.

\section{Theoretical method}
\begin{figure}[!htbp]
    \centering
   \includegraphics[scale=0.3]{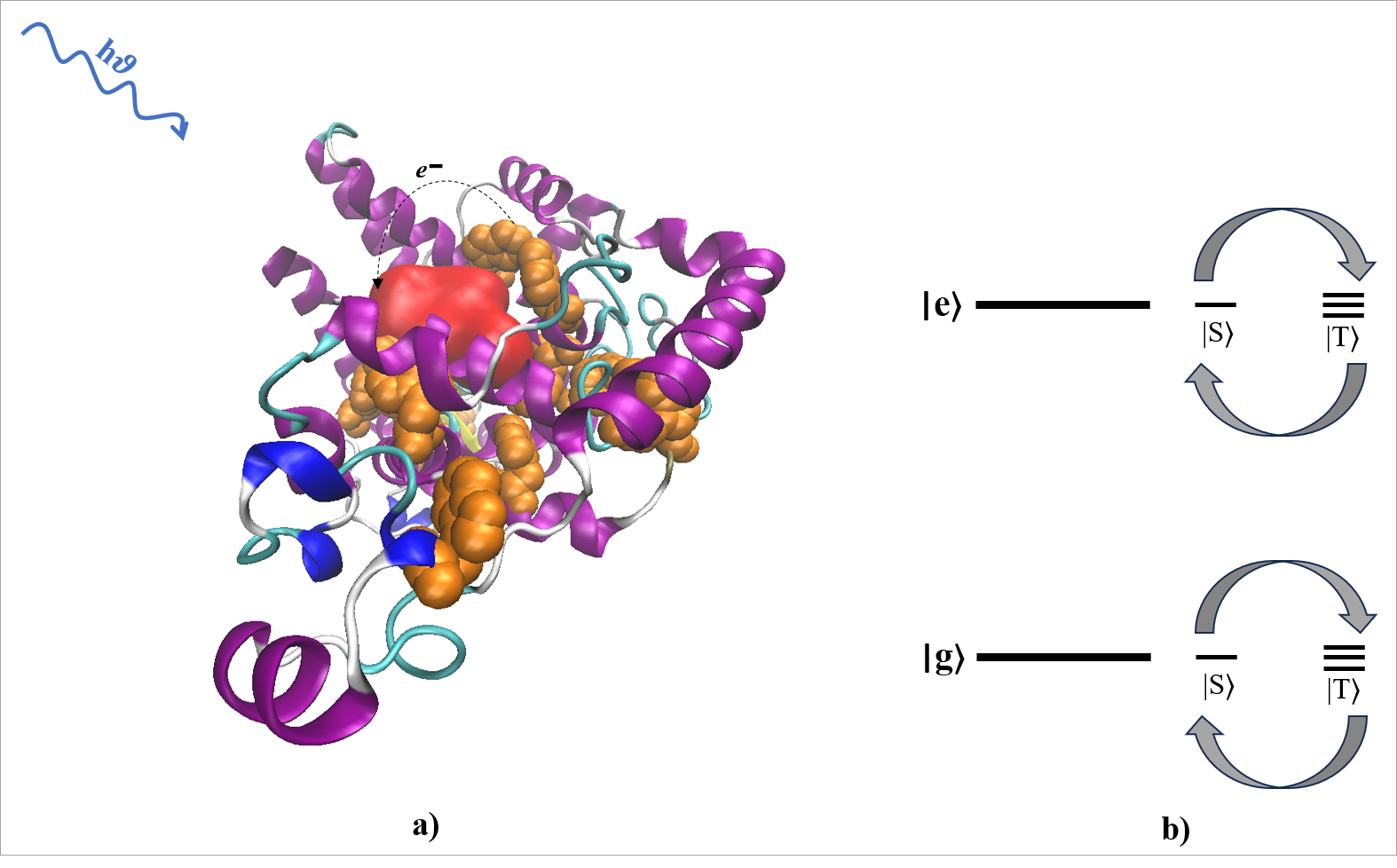}
    \caption{a) Schematic representation of the protein cryptochrome and b) the energy levels of radical pairs consisting of spatial and spin states. Due to hyperfine interaction, radical spin states, engages in singlet to triplet interconversions}
    \label{re_my}
\end{figure}
The absorption of electromagnetic waves, having specific intensity and wavelength, by the flavin adenine dinucleotide (FAD) chromophore can trigger a chain of electron transfer in cryptochrome protein that results in the formation of paired radical intermediates (Fig. \ref{re_my}). Initially, the state of these spin-correlated radical pairs can be either antiparallel to one another (singlet) or parallel (triplet). The spins of the unpaired electrons can interconvert under the influence of factors such as weak magnetic fields (strong magnetic fields can disrupt the mechanism), and it is the general basis of cryptochrome-based magnetoreception.
We begin our analysis by considering the hyperfine interaction, which can play an important role in the spin dynamics and magnetosensitivity of radical pairs. The interaction of the internal magnetic field caused by the magnetic moments of atomic nuclei with the unpaired electron is called the hyperfine interaction \cite{hore2016radical}. The directional asymmetry in the molecular environment is indicated by a diagonal anisotropic hyperfine tensor ($\mathbf{A}= \mathrm{diag} (A_x, A_y, A_z) $) \cite{xu2013estimating}. In the present toy model, following the minimal framework for the radical-pair mechanism described by Aiello et al. \cite{aiello2026physicist} and Woodward \cite{woodward2026low}, we considered only a single spin-$1/2$ nucleus to reduce computational complexity. However, Smith et al. \cite{smith2024optimality} showed that increasing the complexity of a radical-pair system by incorporating additional nuclear spins coupled to the electron spins may enable the system to approach optimal precision. In a separate study, Smith et al. \cite{smith2022observations} reported the counterintuitive observation that electronic coherence can persist in more complex systems. In our model, we focus on the role of the hyperfine interaction in driving interconversion between the singlet and triplet states, neither of which is an eigenstate of the hyperfine Hamiltonian. The effect of system complexity on the precision of the model is left for future work. The Zeeman interaction describes the oscillations in the electron's magnetic moment due to the interaction of an external magnetic field (such as the Earth's magnetic field) with the spin of electrons \cite{hore2016radical}. The intensity of these oscillations depends on the strength of the magnetic field. 

\begin{figure}[!htbp]
    \centering
   \includegraphics[scale=0.4]{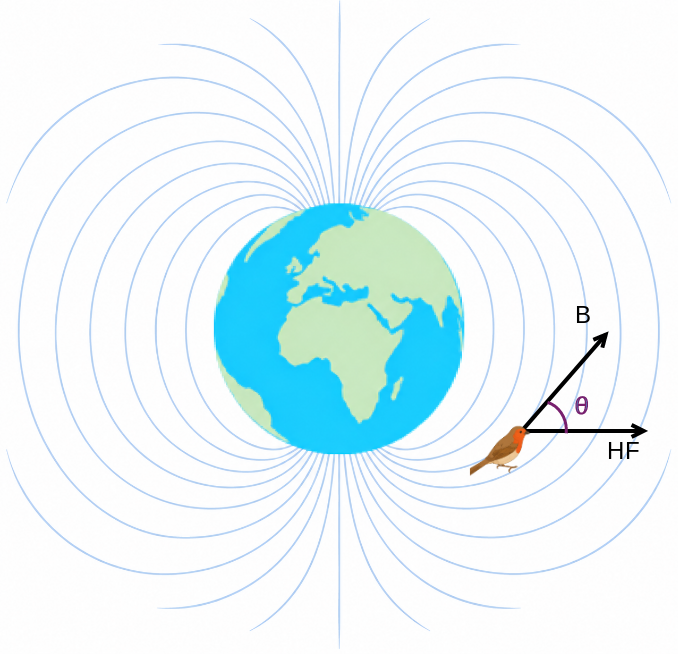}
    \caption{Magnetic field vector inclination angle $\theta$, describes its orientation to the basis of hyperfine vector.}
    \label{coordinate}
\end{figure}

The Earth's magnetic field in the absence of an oscillating field in spherical coordinates system is as follows:
\begin{equation}
\mathbf{B} = B_0 (\sin \theta \cos \phi, \sin \theta \sin \phi, \cos \theta),
\end{equation}
which is represented in Fig. \ref{coordinate}.
To simplify the analysis, assuming axial symmetry in the system, the angle $\phi = 0$ is considered \cite{xu2013estimating}.

In general, the strength of the Zeeman interaction is related to the magnetic field strength and the gyromagnetic ratio ($\gamma$).
Where 
\begin{equation}
  \gamma = \mu_B g_s
\end{equation}
determines the extent to which the electron's magnetic moment couples to the external magnetic field.  $ \mu_B$  is  the Bohr magneton and $g_s$ is the electron spin g-factor ($g_s\approx 2)$.

The time evolution of the considered system is described based on the von Neumann equation.
\begin{equation}
\frac{d}{dt} \rho(t)= -\frac{i}{\hbar} [{H}, \rho]
\label{von}
\end{equation}

where $\rho$ is the density matrix, and $H$ is the Hamiltonian of the system.
Considering the above and in the absence of inter-radical interactions \cite{adams2018open}, the spin Hamiltonian of the radical pair system, which includes hyperfine and Zeeman interactions, is expressed as follows: 

\begin{equation}
{H} = \mathcal{I} \cdot A \cdot {S}_1 + \gamma B \cdot ({S}_1 + {S}_2)
\label{Hamiltonian}
\end{equation}
where $\mathcal{I}$ defines the nuclear spin operator, $S_1$ and $S_2$ are the spin operators of the two unpaired electrons in the radical pair.
The ratio between the strength of the hyperfine and Zeeman interactions has significant effects on the quantum behavior of the radical pair system. Accordingly, if the ratio ($\frac{A_z}{\gamma B_0}$) is small and in the range (0, $2\times10^{-3}$), the Zeeman interaction dominates, and if this ratio exceeds 3, the hyperfine regime will dominate. In conditions where this ratio is in the intermediate range between $2\times10^{-3}$ and 3, neither interaction alone dominates and the effects of both factors are involved in the dynamics of the system \cite{xu2013estimating}.

In this framework we considered $\hbar=1$. $B_0$ represents the magnitude of the Earth's magnetic field and varies between $2.2\times10^{-5}\mathrm{T}$ to $6.7\times10^{-5}\mathrm{T}$. The gyromagnetic ratio $\gamma=0.5\mu_B g_s=1.76\times10^{11}\mathrm{s}^{-1}\mathrm{T}^{-1}$. We added the $1/2$ factor because of using pauli-matrices for spin operators. The implementation of hyperfine effect can vary depend on the effective number and geometrical consideration of nuclei in the system. However, as we discussed before, the most important effect of hyperfine term is the unisotropic balance of hyperfine coefficients between the two spin systems, thus we considered a single spin-$1/2$ nucleus. Hyperfine interaction considered in a regime which approximately is in the order of $10^{-3}$ times of Earth's magnetic field. Accordingly, we considered $A_x=A_y=A_z/2$ and $A_z=8.096 \times 10^3 \mathrm{s}^{-1}$ \cite{xu2013estimating}.

The effects of the interaction between spin and spatial momentum are ignored in most common radical pair models. However, according to the Dirac equation, an electron moving in an electric field experiences an effective magnetic field, which leads to the formation of spin-orbit coupling (SOC). Binhi, V. N., considers the motion of the electron in the electric field as the main cause of the spin change, rather than the magnetic field of the nuclei \cite{binhi2019nonspecific}. This interaction is of the fine-structure type and links the electron’s spatial position to its spin state. SOC is usually stronger than hyperfine interactions and makes the spin evolution dependent on molecular orbital symmetry. Izmaylov et al. included all relevant interaction terms, including spin-spin and spin-orbit interactions, in their model \cite{izmaylov2009relativistic}. They showed that the spin-spin interaction decays with increasing inter-radical separation, whereas the spin-orbit interaction affects singlet-triplet interconversion only at short inter-radical distances. The interaction can play a notable role in magnetosensitivity even in the absence of nuclear spin through entanglement between spatial and spin states. Accordingly, it seems necessary to consider spatial degrees of freedom in modeling radical pairs. Considering the spin-orbit coupling Hamiltonian as follows:
\begin{equation}
H_{SOC}=\zeta_{SOC} L\cdot S 
\label{SOC}
\end{equation}
This term couples the spin angular momentum of the electrons to their orbital angular momentum. Here, $L$ and $S$ denote the orbital and spin angular momentum operators, respectively, and $\zeta_{SOC}$ is the spin--orbit coupling constant. For most organic compounds, the spin--orbit coupling energy is on the order of $0.5$--$5\,\mathrm{meV}$. In the present system, however, the orbital angular momentum is quenched by the molecular field, so the electron spin couples to only a fraction of it. The relevant effective coupling is comparable in magnitude to the anisotropic Zeeman interaction, approximately $1.4\,\mathrm{peV}$. Therefore, we set $\zeta_1=100\,\mathrm{s}^{-1}$ and $\zeta_2=200\,\mathrm{s}^{-1}$ \cite{lambert2013radical}. The difference between these coupling constants introduces anisotropy into the system; however, at larger inter-radical separations, this contribution is negligible compared with the anisotropy induced by the hyperfine interaction.
In this model, the orbital angular momentum is considered as a spin-1 system in a three-dimensional Hilbert space consisting of three sublevels with magnetic quantum numbers $m_l = {-1, 0, +1}$. This corresponds to a three-level orbital model, the lowest-dimensional representation capable of describing electronic transitions with nonzero spin-orbit coupling; a transition between two $m_l=0$ states carries no orbital angular momentum. For each radical, however, two orbital configurations are considered: one corresponding to the orbital angular momentum of the electron in the ground state and the other to that in the excited state. The strength of the coupling is determined by a coupling constant that depends on the electric field experienced by the electron and also on the electronic configuration of the molecule. Based on the magnitude of this constant, the spin-orbit coupling can be classified into one of three regimes: weak, moderate, or strong.
Although the radical pairs in cryptochromes are made of organic molecules, the relatively large structure and the presence of $\Pi$-conjugated systems cause them to have significant SOC even without the presence of heavy elements.
Now, considering the spin-orbit interaction, the Hamiltonian of the system can be written as below to obtain a more accurate description of the quantum dynamics of the radical pair.

\begin{equation}
{H} = \mathcal{I} \cdot A \cdot {S}_1 + \gamma B \cdot ({S}_1 + {S}_2) + \sum_{j=1}^{2} \zeta_j \, {L}_j \cdot {S}_j
\label{FullHamiltonian}
\end{equation}
where $\zeta$ is spin-orbit coupling strength and $j$ refers to each radical. The spin-orbit coupling of organic compounds are really weak considering other effects. Thus in this work we assumed $\zeta=1-2meV$ \cite{winter2017importance}.
In this model, the tensor product of a 2-level nuclear spin by two 2-level electron spins by two 3-level spatial momentum for each electron creates a 72-dimensional Hilbert space.
The initial spin state of the radical pair is considered to be a singlet, defined as follows:
\begin{equation}
\vert\psi_e\rangle=\frac{1}{\sqrt{2}} (\vert 0 1\rangle - \vert 1 0\rangle)
\end{equation}

The initial state of the nuclear spin is considered to be a completely mixed state, which is described by the following density matrix:
\begin{equation}
\rho_I= \frac{1}{2}(\left| \uparrow \right\rangle \left\langle \uparrow \right| + \left| \downarrow \right\rangle \left\langle \downarrow \right|)
\end{equation}

The spatial states are considered as a two-level system that transitions from the ground state $ |g\rangle $ with $L_z=-1$ to the excited state $|e\rangle $ with $L_z=0$. In this model, for compatibility with real atomic states, we considered a three dimensional space for the angular momentum but only two of the three levels are being used.
The initial spatial state of the system is defined as follows:
\begin{equation}
\label{spatial}
\vert\psi_{spatial}\rangle= \frac{1}{2}(\vert g\rangle _1+ \vert e\rangle _1 )(\vert g\rangle _2+ \vert e\rangle _2 )
\end{equation}
we considered our initial state entangled in its spin states (singlet) and separated in spatial states (Fig \ref{re_my}). However, during the time evolution of the system, both the spatial and spin components become entangled. Accordingly, the density matrix of the initial state is given by
\begin{align}
\rho_i &= \rho_I \otimes \rho_{\mathrm{spatial}} \otimes \rho_e, \nonumber \\
\rho_{\mathrm{spatial}} &= |\psi_{\mathrm{spatial}}\rangle\langle\psi_{\mathrm{spatial}}|,
\qquad \rho_e = |\psi_e\rangle\langle\psi_e|.
\end{align}
Our main goal in considering the interaction between electron spin dynamics and its spatial dynamics is to determine the biophysical mechanism for detecting or converting the Earth's magnetic field information into understandable signals through the electric dipole moment.

We expect this dipole moment to be sensitive to the properties of the external magnetic field. On the other hand, the electric field resulting from the dipole moment may be able to interact with surrounding biological structures and thus establish a pathway for transmitting the necessary information for magnetic navigation.
The dipole moment operator for the radical pair system is: 
\begin{equation}
\mathbf{p} = \sum_i \mathbf{e }\, \mathbf{r_i}
\end{equation}
Where $\mathbf{e}$  is the elementary charge , and $\mathbf{r}$ is the position operator.
The vector operator $\mathbf{r}$ is expressed as:

\begin{equation}
\mathbf{r} = \hat{x} \, {\imath} + \hat{y} \, {\jmath} + \hat{z} \, {k}
\end{equation}
where $\hat{x}$, $\hat{y}$, and $\hat{z}$ are the position operators along the x, y, and z directions, and $i,j$ and $k$ are the unit vectors of Cartesian coordinate system, respectively.

Generally, the position operator acts on a continuous, infinite-dimensional Hilbert space, whereas the radical-pair system is modeled in a finite-dimensional, spin-based Hilbert space. Therefore, continuous position operators cannot be used directly. Here, we adapt the definition of the position operator from the harmonic-oscillator construction to a two-level system and apply the rotating-wave approximation. Specifically, we construct the position operator along the direction of the radical-pair dipole moment ($\mathbf{r}$) and define the corresponding ladder operators in terms of the ground state $|g\rangle$ and excited state $|e\rangle$ as
\begin{align}
    \sigma^+_{(r)}=|e\rangle\langle g|, \qquad \sigma^-_{(r)}=|g\rangle\langle e|.
\end{align}
Assuming that the radical-pair dipole moment is oriented along the $\mathbf{r}$ axis, the position operator is given by
\begin{align}
    \hat{\mathbf{r}} = \sqrt{\frac{\hbar}{2m\omega}} \left( \hat{\sigma}^+ + \hat{\sigma}^- \right).
\end{align}

To determine the coefficient of the position operator, we could use the bare-electron mass, $m_e=9.11\times 10^{-31}\,\mathrm{kg}$, as a standard effective mass for electron transfer in a protein and take the energy gap to be on the order of $1\,\mathrm{eV}$. These scales yield $r_0=\sqrt{\hbar/(2m_e\omega)}\approx 2\,\text{\AA}$. However, to implement the operator more realistically, we use the reported radical separation of $15.4\,\text{\AA}$ \cite{timmer2023tracking}. Taking into account the electron charge and the relative positions of the two radicals, we obtain
\begin{align}
    \mathbf{P} = -1.23\times10^{-28}\left[(\sigma^+_1 + \sigma^-_1)-(\sigma^+_2 + \sigma^-_2)\right].
\end{align}
The matrix representation of this operator is
\begin{align}
    \mathbf{P} = -1.23\times10^{-28}\left[
    \begin{pmatrix}
    0 & 1 \\
    1 & 0 \\    
    \end{pmatrix}\otimes I - I\otimes
    \begin{pmatrix}
    0 & 1 \\
    1 & 0 \\ 
    \end{pmatrix}\right],
\end{align}
where $I$ denotes the identity operator. Our model is intentionally minimal. To apply this framework to a realistic biomolecular system, the minimal Hamiltonian would be replaced by a system-specific spin Hamiltonian that includes hyperfine, Zeeman, spin–orbit, exchange, and dipolar interactions. In a real molecule, the hyperfine and $g$-tensors are anisotropic and must be determined from EPR spectroscopy or DFT calculations. The single-nucleus term in our model would then be extended to include all relevant nuclear spins, or replaced by an effective spin bath if the remaining nuclei are treated as an environment that causes relaxation and dephasing. Spin–orbit coupling could likewise be included explicitly or incorporated into effective $g$-tensor and zero-field-splitting parameters.

Indeed, so far we have considered the radical pair as a closed and ideal system. But to describe the realistic behavior of a radical pair, we must regard environmental effects. Because radical pairs of cryptochrome protein are not isolated and interact with surrounding biomolecules and solvent at high physiological temperatures. These interactions cause decay of system coherence and destroy quantum properties of biological structure. Here, we have modeled the radical pair as an open quantum system that interacts with the environment through the Lindblad master equation \cite {breuer2002theory} as follows:
\begin{equation}
\frac{d}{dt} \rho_s = -\frac{i}{\hbar} [H_s, \rho_s] + \mathcal{L}(\rho_s)
\end{equation}
Where ${L}$ is the Lindblad superoperator, which is expressed as follows:
\begin{equation}
\mathcal{L}(\rho_s) = \Gamma \sum_i \left(A_i\rho_s A_i^\dagger-\frac{1}{2}\{A_i^\dagger A_i,\rho_s\}  \right)
 \label{lin}
\end{equation}
 Here, $\Gamma$ is the dissipation rate, and $A$ denotes the jump operators acting on the spatial states. Specifically, $A_1=\sigma_1^-\otimes I_2$ and $A_2=I_1\otimes\sigma_2^-$, where $I$ is the identity operator; $A_3$ and $A_4$ are the corresponding conjugate operators, as defined in Eq.~\eqref{spatial}. This formulation provides a more realistic description because biological systems couple to the environment primarily through their positional degrees of freedom. Because the spin and spatial states are coupled, the electron spins also interact indirectly with the environment. The dissipation rate, $\Gamma$, depends on the environmental temperature and the system's oscillation frequency. If the environment is modeled as a thermal bath of harmonic oscillators with an Ohmic spectral density, the temperature dependence of $\Gamma$ is given by $\Gamma=\gamma_0^2\Omega\bar{n} r^2/(1+r^2)$ \cite{gorini1976completely, maniscalco2004lindblad}. Here, $\gamma_0$ is the system-environment coupling strength, $\Omega$ is the system's oscillation frequency, and $\bar{n}$ is the mean occupation number of the oscillators in the thermal bath. The latter depends on the environmental temperature, $T$, and is given by $\bar{n}=(e^{\Omega/kT}-1)^{-1}$. The parameter $r$ characterizes the distribution of bath-oscillator frequencies relative to the cutoff frequency. In biological environments, $\Gamma$ ranges from $10^6$ to $10^8$ \cite{vaziri2010quantum}.

We note that this model does not include spin-selective radical-pair recombination. Therefore, the long-time behavior calculated later in this paper should not be interpreted as the conventional reaction-yield response of a finite-lifetime radical pair.


\section{Numerical Results and Discussion}
An adaptive Adams-Bashforth-Moulton linear multistep method, with relative and absolute tolerances of $10^{-8}$, was used to solve the von Neumann equation, Eq.~\eqref{von}, and determine the time evolution of the radical-pair system. Convergence was verified using tighter tolerances and an independent explicit Runge-Kutta integrator. We considered using numerical approach due to the considerable complexity of the system and the size of the Hilbert space, which incorporates multiple quantum degrees of freedom. This structure enables the model to coherently capture spin dynamics, spatial asymmetry, and spin–orbit interactions as the key elements underlying the emergence and modulation of the electrical dipole moment under external magnetic field influence. To estimate the electric dipole moment using the radical-pair generation rate reported in Ref.~\cite{solov2012reaction}, we assumed a population of $3\times10^7$ radical pairs. This is due to the generation rate of the radical pairs and their lifetime which is about $10^6\mathbf{s}$. For a single radical-pair, the values simply can divide by the number of radical-pairs. Also, we converted the dipole-moment value from $\mathbf{C.m}$ to Debye which is approximately $3.34\times10^{-30}\mathbf{C.m}$. A complete table of the parameters used in our calculations is represented in TABLE \ref{tab:parameters}

\begin{figure}[!htbp]
    \centering
    \subfloat[\label{fig:time-short}]{\includegraphics[width=0.48\linewidth]{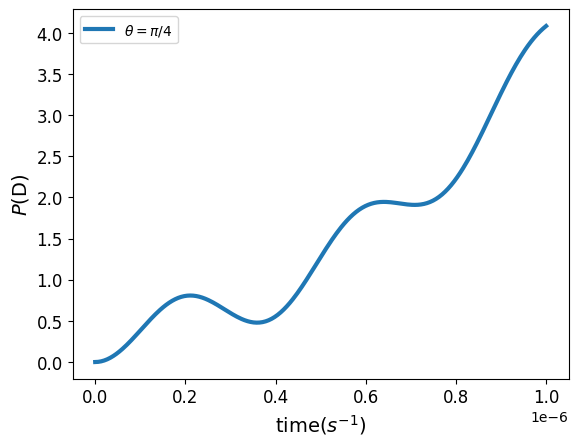}}
    \hfill
    \subfloat[\label{fig:time-long}]{\includegraphics[width=0.48\linewidth]{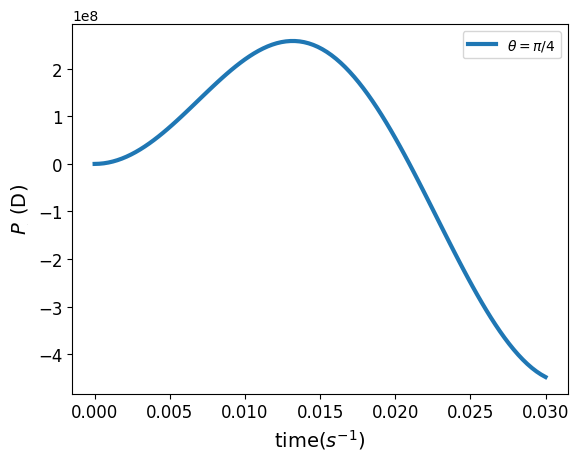}}
    \caption{Time evolution of the electric dipole moment $P$ at $\theta=\pi/4$. (a) Evolution of $P$ on the microsecond timescale. (b) The periodic behavior of $P$ on a timescale of approximately $10^{-2}\,\mathrm{s}$ confirms the mathematical consistency of the model. However, in the absence of dissipative effects, this long-time behavior is not physically meaningful.}
    \label{fig:two_subfigures}
\end{figure}

As shown in Fig.~\ref{fig:two_subfigures}, the expectation value of the dipole moment, $P$, increases on the microsecond timescale without showing an apparent decline. However, this expectation value must remain bounded and cannot increase indefinitely. We therefore extended the calculation to $10^{-2}\,\mathrm{s}$ to demonstrate its periodic behavior and upper bound. These long-time calculations are included only to verify the mathematical consistency of the solution and are not intended to represent physically meaningful system behavior. In a realistic long-time regime, dissipative effects damp the response to lower values, as discussed in the dissipative-dynamics section.

Since we understood the general behaviour of the system, we studied the evolution of the  expectation value of the electrical dipole moment through time, in a variety of inclination angle of magnetic field lines ($\theta$). As shown in Fig.~\ref{Dipole_vs_time}, the time evolution of the expectation value of the dipole moment, $P$, depends strongly on the inclination angle. Moreover, Fig.~\ref{labeled_ab} indicates that this sensitivity varies periodically with the inclination angle.

\begin{center}
\begin{minipage}{\linewidth}
    \centering
    \captionof{table}{Physical Parameters and Their Values}
    \label{tab:parameters}
    \footnotesize
    \begin{tabular}{@{}lcl@{}}
        \toprule
        \textbf{Parameter} & \textbf{Symbol} & \textbf{Value} \\
        \midrule
        Reduced Planck constant & $\hbar$ & $1$ \\
        Earth's magnetic field & $B_0$ & $(2.2$--$6.7)\times 10^{-5}\,\mathrm{T}$ \\
        Gyromagnetic ratio & $\gamma$ & $1.76\times10^{11}\,\mathrm{s^{-1}T^{-1}}$ \\
        Hyperfine coefficient & $2A_x=2A_y=A_z$ & $8.096 \times 10^{3}\,\mathrm{s^{-1}}$ \\
        Spin-Orbit coefficient & $\xi_1$, $\xi_2$ & $100, 200\,\mathrm{s^{-1}}$ \\
        Dissipation rate & $\Gamma$ & $1\times10^6\,\mathrm{s^{-1}}$ \\
        Thermal bath temperature & $T$ & $300\,\mathrm{K}$ \\
        \bottomrule
    \end{tabular}
\end{minipage}
\end{center}

\begin{center}
\begin{minipage}{\linewidth}
    \centering
    \includegraphics[scale=0.5]{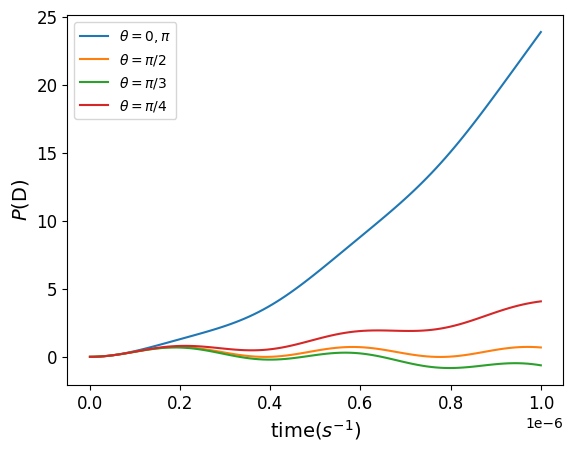}
    \captionof{figure}{Time evolution of the electric dipole moment $P$ over one microsecond for different magnetic field inclination angles, $\theta$. The expectation value of the electric dipole moment varies with the inclination angle.}
    \label{Dipole_vs_time}
\end{minipage}
\end{center}
In general, as a bird moves, the orientation and magnitude of the magnetic field relative to the bird may change, thereby affecting the electric dipole moment. This change could then be interpreted as an orientation signal. Extensive research has examined the sensitivity of the quantum needle to magnetic field orientation in chemical magnetoreception. Previous studies have shown that interradical modulations induced by molecular motion can bring the angular precision of magnetic field sensing closer to the quantum limit \cite{smith2026interradical} and that hyperfine coupling can be optimized to improve directional precision in chemical magnetoreception \cite{chen2024identifying}. In this study, we performed computational modeling to elucidate the effects of the magnetic field's inclination angle, $\theta$, and magnitude on the quantum dynamics of radical pairs.

Firstly, we investigated the changes in the electric dipole moment with respect to the magnetic field’s inclination angle. The analysis was carried out in the order of microseconds, and the system was considered to be closed and coherent during this time. It should be noted that the effects of decoherence under physiological temperature have been investigated in Subsection \ref{dissipative}. Fig. \ref{labeled_ab}a shows the variation of the magnetic field effect (MFE)
\begin{align}
    \text{MFE}(B) = \frac{P(0)-P(B)}{P(0)}
\end{align} 
for different inclination angles of the magnetic field. The results reveal a notable and periodic dependence on $\theta$. Furthermore, identical values at $\theta = 0$, $\pi$, and $2\pi$ reflect that although the system is sensitive to the inclination angle, it does not detect polarity of the magnetic field.This finding is consistent with reported observations \cite{mcfadden2016life}.
\begin{center}
    \centering
    \includegraphics[scale=0.5]{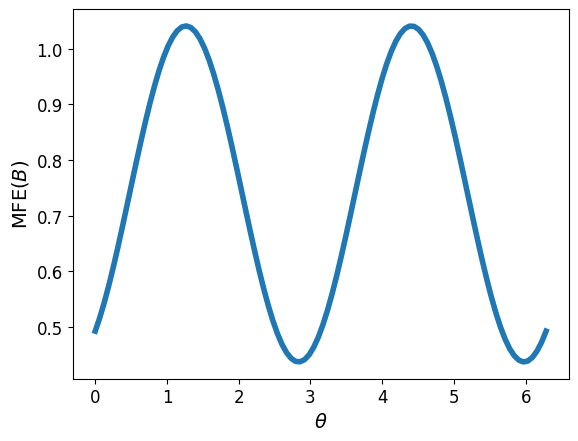}
    \captionof{figure}{Variations of the magnetic field effect (MFE) with respect to the magnetic field inclination angle $\theta$ at t = $1\mu s$. the dipole moment shows a clear periodic dependence on $\theta$, with distinct peaks indicating the angular sensitivity of the system.}
    \label{labeled_ab}
\end{center}

In general, radical-pair generation occurs on a picosecond timescale \cite{solov2012reaction}. By contrast, the response and signal-processing times in these structures may be much longer than the quantum coherence time, but given the average flight speed of the bird, it is expected that there will be no noticeable change in its position during this time frame. Consequently, it can be assumed that the magnetic field experienced by the radical pair system remains constant during this time, and the directional information carried by the electric dipole moment can be used to navigate the bird. It is expected that the radical pair in cryptochrome can only be considered a magnetic receptor if it is capable of participating in signal transduction processes \cite{gortemaker2022direct}. The electrical signal resulting from the electric dipole moment related to radical pair may be transmitted through changes in the current of ion channels and cell transmembrane potential or changes in the conformation of cryptochrome-interacting proteins such as G-proteins \cite{gortemaker2022direct} and ultimately interpreted in a specific part of the brain.

Then, following prior evidence \cite{hore2019upper} that radical pair dynamics in cryptochrome can be influenced by magnetic field strength, we evaluated the changes in the electric dipole moment with respect to the intensity of the external magnetic field for three different angles at one microsecond. As can be seen in Fig. \ref{Dipole_vs_B}, at lower intensities, especially in the geomagnetic range (approximately $25$–$65\mu T$), there is an oscillatory and nonlinear dependence on $B_0$, such that small variations in the field intensity lead to significant changes in the dipole moment amplitude. At intensities beyond the geomagnetic window, a gradual decrease in the dipole moment amplitude is revealed, and as a result, the system is less responsive to changes in the field strength.
\begin{center}
    \centering
    \includegraphics[scale=0.5]{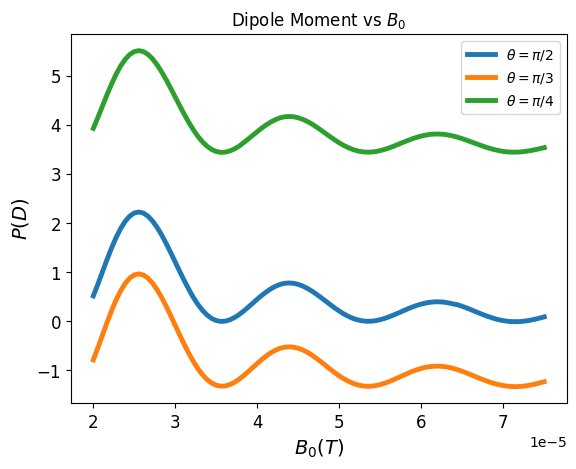}
    \captionof{figure}{Variation of the electric dipole moment with magnetic field strength at one microsecond. The dipole moment exhibits oscillations of decreasing amplitude as the magnetic field strength increases.}
    \label{Dipole_vs_B}
\end{center}

We generated a three-dimensional plot to illustrate how the electric dipole moment, responds to simultaneous changes in the magnetic-field inclination angle, $\theta$, and strength, $B_0$. Because both properties can vary with latitude, Fig.~\ref{Dipole Moment vs R} shows that $P$ depends periodically on $\theta$ and is also sensitive to $B_0$.
Together, these dependencies suggest that the mechanism may provide information beyond direction, including cues related to location. Thus, radical-pair-based magnetoreception may provide directional and intensity cues relevant to magnetic sensing. Our findings support the possibility that the avian sensory system interprets changes in the electric dipole moment as such cues.
\begin{center}
    \centering
    \includegraphics[scale=0.5]{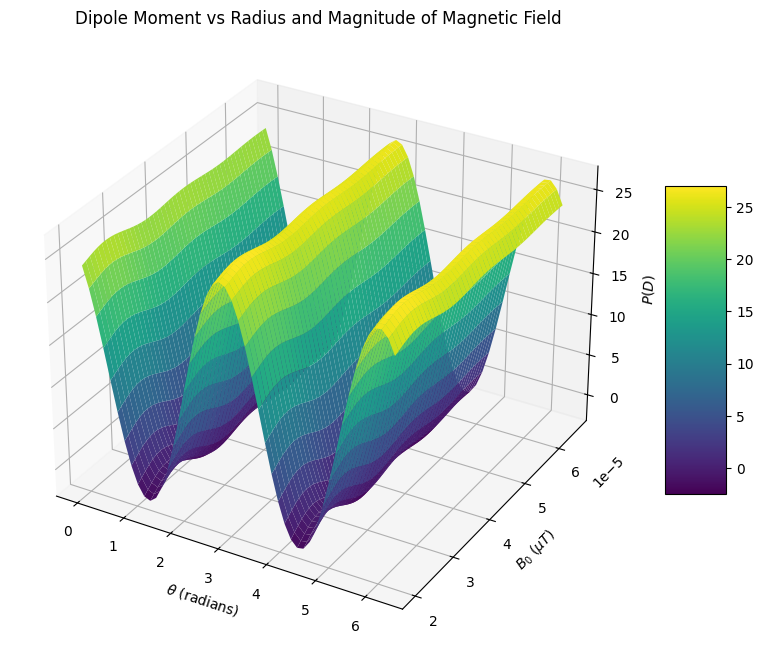}
    \captionof{figure}{3D plot representation of variation of dipole moment with magnetic field angle and strength at one microsecond}
    \label{Dipole Moment vs R}
\end{center}

\subsection{Dissipative effect}
\label{dissipative}
\begin{figure}[h]
    \centering
    \includegraphics[scale=0.4]{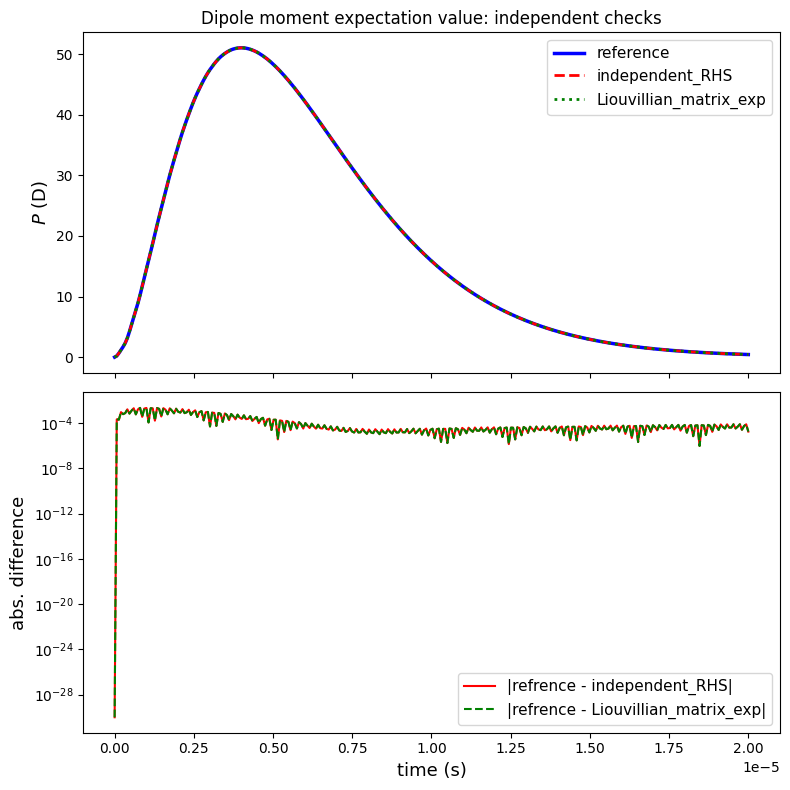}
    \caption{Top: Time evolution of the electric dipole moment calculated using three independent methods for solving the Lindblad equation. Bottom: Absolute differences between the results on a logarithmic scale, demonstrating agreement within the integration tolerance.}
    \label{fig:three_methods}
\end{figure}

We now study the time evolution of the radical-pair system under dissipative effects using the Lindblad master equation, Eq.~\eqref{lin}. Noting that the physiological temperature plays a key role in the  dynamics of open quantum systems by changing the rate of dissipative processes, this effect is implicit in the $\Gamma $ parameter of the master equation \cite{soltanmanesh2019clausius}. In the following, the dissipation role in the time evolution of the x-component of the electric dipole moment for various magnetic field inclination angles was investigated. 
\begin{figure}[!htbp]
    \centering
    \subfloat[\label{fig:D_diss_pi}]{\includegraphics[width=0.48\linewidth]{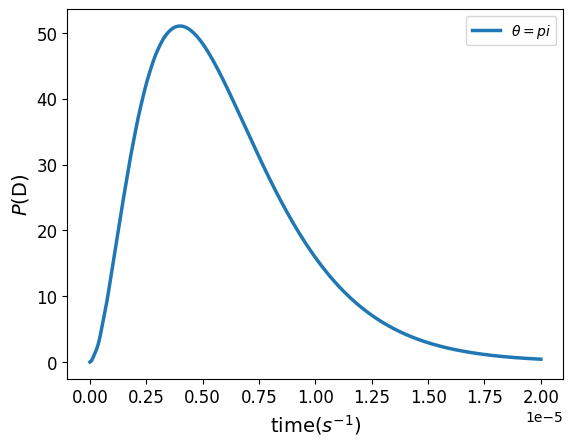}}
    \hfill
    \subfloat[\label{fig:D_diss_pi_2}]{\includegraphics[width=0.48\linewidth]{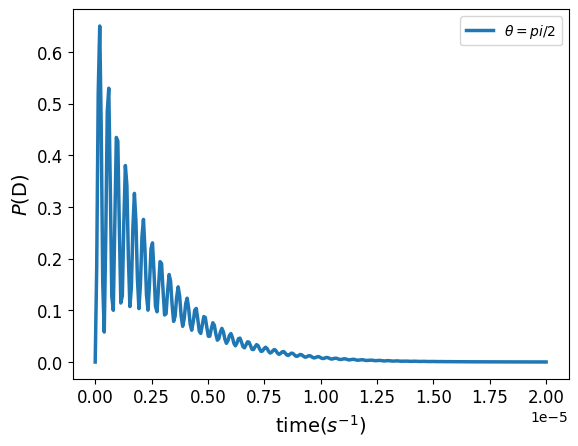}}
    \caption{Time evolution of the electric dipole moment for different magnetic field inclination angles $\theta = \pi$ (a) and $\pi/2$ (b) under the influence of environmental dissipation}
    \label{Time evolution dissipation}
\end{figure}
We validated the adaptive Lindblad integrator used throughout this work against an independent numerical integration of the Lindblad equation and direct matrix exponentiation of the Liouvillian. The three methods agree within the integrator tolerance over the entire time interval. Fig.~\ref{fig:three_methods} compares the results at an inclination angle of $\theta=\pi$ and shows the maximum differences among the methods. Positivity and trace preservation were also verified throughout the time interval. For all three methods, the maximum deviation from unit trace, $|\operatorname{tr}(\rho)-1|$, was of the order of $10^{-15}$.

As shown in Fig.~\ref{Time evolution dissipation}, the system’s behavior under environmental dissipation depends on the inclination angle of the external magnetic field. At $\theta=\pi$, the dipole moment initially increases, indicating strong coherent dynamics, and then decays exponentially because of decoherence induced by environmental interactions. At $\theta=\pi/2$, damped oscillations emerge, reflecting the dissipative effect, whereas at $\theta=\pi/3$, more irregular oscillations indicate a transition to a less coherent state.

\begin{figure}[!htbp]
    \centering
    \includegraphics[scale=0.5]{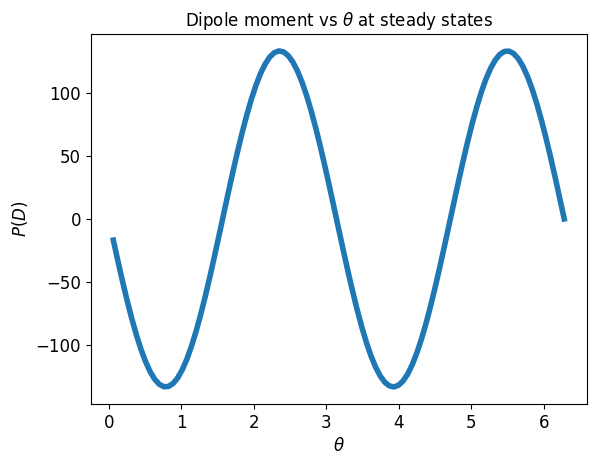}
    \caption{Variation of the electric dipole moment $P$ with the magnetic-field inclination angle $\theta$ in the steady-state regime, for an abstract model in which the spin and orbital degrees of freedom are continuously present.}
    \label{steady-state}
\end{figure}

It is anticipated that the dissipative effect induced by thermal noise causes a decrease in the quantum behavior of biological systems over time, but according to Fig.~\ref{steady-state}, it can be seen that when the radical pair system is subjected to dissipative effects and the system meets a stable state (steady-state regime), the changes in the electric dipole moment ($P$) are still angle-dependent.
\section{Concluding remarks}
In this study, we attempted to gain a more detailed understanding of the behavior of radical pairs under the influence of a magnetic field using quantum modeling. The model incorporates spatial degrees of freedom through the spin-orbit coupling term in the Hamiltonian. We computationally investigated how a static magnetic field affects the electric dipole moment of the radical pair. Our results show that the dipole moment depends on both the inclination angle and the magnitude of the applied magnetic field.

Moreover, simultaneous changes in the field's inclination and strength suggest that the system may provide information about both direction and location. The modeled radical pair also retains magnetic sensitivity when dissipation and decoherence are included. The steady-state results show that the electric dipole moment remains dependent on the magnetic-field inclination angle, indicating that the system retains a field-dependent response despite decoherence. The model therefore suggests a possible route by which electric dipoles could respond to magnetic fields even under strong dissipation, which is a valuable achievement not only in the field of magnetoreception but also in quantum biology. More speculatively, if such effects are experimentally confirmed, they could inform the design of bio-inspired or synthetic spin-chemical magnetic sensors and motivate studies of weak-field biological responses.

\section*{Authors' contributions}
A.R.B and M.S proposed the research problem; A.S modeled the system and mathematically formulated the problem and computed the results; S.A, N. H, S. R and A.B analyzed the data; M.S wrote the first draft, and all authors edited the final manuscript. 


We refer to Wiltschko and Wiltschko's (2014) study, ``Sensing magnetic directions in birds: radical pair processes involving cryptochrome'' (\textit{Biosensors} 4(3), 221--242), which discusses how the magnitude of an external magnetic field can affect the avian magnetosensory system. Because this topic is beyond the scope of our manuscript, we have removed the sentence.

\section*{Funding}
This research did not receive any specific grant from funding agencies in the public, commercial, or not-for-profit sectors.

\section*{Data Availability }
All data used in this study are publicly available on GitHub at: \url{https://github.com/Alimeril/radical_pair_supplementary_data}.

\end{document}